\documentclass[fleqn,10pt]{wlscirep}
\usepackage{amsmath, mathrsfs, latexsym, amssymb, color, graphicx, comment, hyperref, scrhack}
\usepackage{multirow}
\usepackage{lmodern}
\usepackage{physics}
\usepackage[utf8]{inputenc}
\usepackage[T1]{fontenc}

\title{Superconducting phase transitions in disordered NbTiN films}

\author[1,2]{M.\,V.\,Burdastyh}
\author[1,3]{S.\,V.\,Postolova}
\author[4]{T. Proslier}
\author[3]{S.~S.~Ustavshikov}
\author[3]{A. V. Antonov}
\author[5]{V.\,M.\,Vinokur}
\author[1,2]{A.\,Yu.\,Mironov}
\affil[1]{A.\,V.\,Rzhanov Institute of Semiconductor Physics SB RAS, 13 Lavrentjev Avenue, Novosibirsk 630090, Russia}
\affil[2]{Novosibirsk State University, Pirogova str. 2, Novosibirsk 630090, Russia}
\affil[3]{Institute for Physics of Microstructures RAS, GSP-105, Nizhny Novgorod 603950, Russia}
\affil[4]{Institut de recherches sur les lois fundamentales de l'univers, Commissariat de l'\'{e}nergie atomique et aux \'{e}nergies renouvelables-Saclay, Gif-sur-Yvette, France}
\affil[5]{Materials Science Division, Argonne National Laboratory, 9700 S. Cass Ave, Argonne, IL 60439, USA}

\newcommand{\rs}{\rm\scriptscriptstyle}

\begin{abstract}
The suppression of superconductivity in disordered systems is a fundamental problem of condensed matter physics.
Here we investigate the superconducting niobium-titanium-nitride (Nb$_{1-x}$Ti$_x$N) thin films grown by atomic layer deposition (ALD) where  disorder is controlled by the slight tuning of the ALD process parameters.
We observe the smooth crossover from the disorder-driven superconductor-normal metal transition (often reffered to as fermionic mechanism) to the case where bosonic mechanism dominates and increasing disorder leads to formation of metal with Cooper pairing.
We show that, in 'moderately' disordered films, the transition to zero-resistance state occurs in a full agreement with the conventional theories of superconducting fluctuations and Berezinskii-Kosterlitz-Thouless transition. However, the 'critically' disordered films violate this accord showing low-temperature features possibly indicating the Bose metal phase.
We show that it is the interrelation between film's sheet resistance in the maximum, $R_\mathrm{max}$, of the resistive curve $R_{\Box}(T)$ and $R_\mathrm{q}=h/4e^2$ that distinguishes between these two behaviors.
We reveal the characteristic features in magnetoresistance of the 'critically' disordered films with $R_\mathrm{max} > R_\mathrm{q}$.
\end{abstract}
\begin{document}

\flushbottom
\maketitle

\thispagestyle{empty}
%%%%%%%%%%%%%%%%%%%%%%%%%%%%%%%%%%%%%%%%%%%%%%%%%
\vspace{-1cm}
\section*{Introduction}
The superconductor-insulator transitions (SIT) observed in a wealth of various systems~(\cite{Gantmaher2010, Goldman2015}), are  usually divided in  two different categories depending on the assumed mechanism of superconductivity suppression --- fermionic~\cite{Finkelstein} or bosonic~\cite{Gold1986, Fisher1990}.
%There are two main  mechanisms of suppression the superconductivity in thin disordered films~\cite{ Gantmaher2010, Goldman2015} --- fermionic~\cite{Finkelstein} and bosonic~\cite{Gold1986, Fisher1990}.
The fermionic scenario proposes that disorder destroys the Cooper pairs.
%and the subsequent localization of the normal electrons. %In this case, the transition from the superconducting to the insulating phase is a two-stage process.
In this case, as the normal state sheet resistance $R_{\Box _\mathrm{N}}$  increases, both superconducting critical temperature $T_\mathrm{c}$ (where the life-time of fluctuating Cooper pairs diverges) and $T_\mathrm{BKT}$ temperature (where the global phase coherence establishes in a system) decrease down to zero and then the superconductor-metal transition occurs due to the complete disappearance of the Cooper pairs, and the suppression of $T_\mathrm{c}$ with $R_{\Box _\mathrm{N}}$ is described by Finkel'stein formula~\cite{Finkelstein}.
%Further, there is a wide region $R_{\Box}$ of metallic states, in which the resistance saturates at non-zero value with cooling down. Finally, the metal-insulator transition occurs for higher $R_{\Box}$.
Such a behaviour is observed, for example, in thin films of niobium nitride NbN \cite{Yong2013, Makise2015}.
Within the bosonic scenario, increasing disorder localizes  Cooper pairs, which exist in the insulating state. Hence, as $R_{\Box _\mathrm{N}}$  increases, $T_\mathrm{c}$ decreases weakly  and $T_\mathrm{BKT}$ decreases down to zero. In this scenario the SIT occurs through a single point of the metal phase with the critical sheet resistance $R_\mathrm{q} =h/4e^2 = 6.45$ k$\Omega$~\cite{Fisher1990}. Experimentally, the critical sheet resistance  is not universal and varies in the range between approximately $R_\mathrm{q}/2$ and $3R_\mathrm{q}$.
The direct disorder-driven SIT is observed, for example, in  indium oxide InO$_{x}$~\cite{Hebard1990, Shahar1992, Gantmakher2000}
%, beryllium Be \cite{Bielejec2001}
and titanium nitride TiN~\cite{Hadacek2004, Baturina2007}  films.
%and recently in niobium-titanium nitride NbTiN \cite{Burdastyh2017}.
At present, the interrelation between these mechanisms as well as the nature of the possible intermediate Bose metal~\cite{Das1999, Phillips2003, BoseArxiv2018, BurmSF} still are not completely understood and are intensively debated~\cite{Goldman2015, BoseShahar2019}.

%Here we report the detailed study of the superconductivity suppression in thin Nb$_{1-x}$Ti$_x$N films on approach to SIT.
Here we examine the approach to the disorder-driven SIT in a  few sets of Nb$_{1-x}$Ti$_x$N films grown by ALD technique. The difference between sets is in the slight shift of ALD  parameters: the fraction $x$ and the temperature of deposition $T_\mathrm{ALD}$.
We show that with $x$ increasing and $T_\mathrm{ALD}$ decreasing the smooth crossover occurs from fermionic mechanism of superconductivity suppression to the case when both bosonic and fermionic mechanisms are involved.
We show that superconducting NbTiN films fall into two categories depending on a ratio between $R_\mathrm{q}$ and $R_\mathrm{max}$, the sheet resistance in the maximum of the $R_{\Box}(T)$ curve.
We reveal the intrinsic difference  between films with $R_\mathrm{max}/R_\mathrm{q}$ smaller and greater than unity through quantitatively  distinct behaviours of the magnetoresistance.
%Possible manifestations of the Bose metal phase in samples with $R_{\Box}_\mathrm{max}/R_\mathrm{q} >1 $ are discussed.
%Superconducting films with $R_\mathrm{max} > R_\mathrm{q}$ do not experience the BKT transition to zero-resistive state.
%Films with $R__{\Box}\mathrm{max} < R_\mathrm{q}$ exhibit thermodynamic superconducting transition in agreement with theory of two-dimensional superconductors.
%is in agreement with the conventional theories of superconducting fluctuations and Berezinskii-Kosterlitz-Thouless transition
%both  $T_\mathrm{BKT}$ and $T_\mathrm{c}$ decrease with disorder in accord with fermionic scenarion of SIT.
%The films with $R_\mathrm{max} > R_\mathrm{q}$ show the substantial resistance drop and corresponding $T_\mathrm{c}$ decreases with $R_{\Box}$ but slower then it should in fermionic approach.

%In the past few decades, the study of the nature of the quantum superconductor---insulator transition~(SIT) in thin disordered films has attracted much %attention~\cite{Goldman, Gantmaher2010}. A pioneering work ~\cite{Shalnikov1938} was the beginning of extensive research on ~ SIT. It was the first to %demonstrate the suppression of the critical transition temperature to the superconducting state, ~ $T_c$, while reducing the sample thickness. Later, it was shown %that the value ~ $T_c$ correlates best with the resistance of the square section of the film $R_{sq}$ --- resistance per square \cite{Strongin1970}.
\vspace{-0.3cm}
\section*{Samples preparation and characterization}
To grow suitable NbTiN films, we employ the atomic layer deposition (ALD) technique based on sequential surface reaction step-by-step film growth.
%The fabrication technique is described in detail in the Supplemental Material.
This highly controllable process provides superior thickness and stoichiometric uniformity and an atomically smooth surface\,\cite{Lim:2003, Driessen:2012} as compared to chemical vapor deposition, the standard technique used to grow NbTiN films\cite{Makise:2016}. We used NbCl$_5$, TiCl$_4$, and NH$_3$ as gaseous reactants; the stoichiometry was tuned by varying the ratio of TiCl$_4$/NbCl$_5$ cycles during growth\,\cite{Proslier:2011}. The superconducting properties of these ultrathin NbTiN films were optimized by utilizing AlN buffer layers grown on top of the Si substrate\,\cite{Shiino:2010}. All films have a fine-dispersed polycrystalline structure~\cite{Mironov2018} with the average crystallite size being $\approx 5$\,nm.

Three sets of Nb$_{1-x}$Ti$_x$N films films are grown varying  deposition temperature $T_\mathrm{ALD}$ and fraction of Ti $x$. For Set-1 $T_\mathrm{ALD}=$450\,$^0$C, and  $T_\mathrm{ALD}=$350\,$^0$C for Set-2 and Set-3. The Ti fraction $x=0.3$  in Set-1 and Set-2 and $x=0.33$ in Set-3.
Films within single Set are grown varying the number of ALD cycles, that provides films of different thickness $d$. The parameters of  samples are given in the Table.
The Hall carrier density $n$ (see SI) appears to be approximately the same regardless of the disorder $n \sim 10^{22}$\,cm$^{-3}$ which is one order smaller that in NbTiN films examined in~\cite{NTN_Cloud}.

%%%%%%%%%%%%%%%%%%%%%%%%%%%%%%%%%%%%%%%%%%%%%%%%%%%%%%%%%%%%%%%%%%%%%%
\begin{figure*}[t!]
\centering
\includegraphics[width=0.95\linewidth]{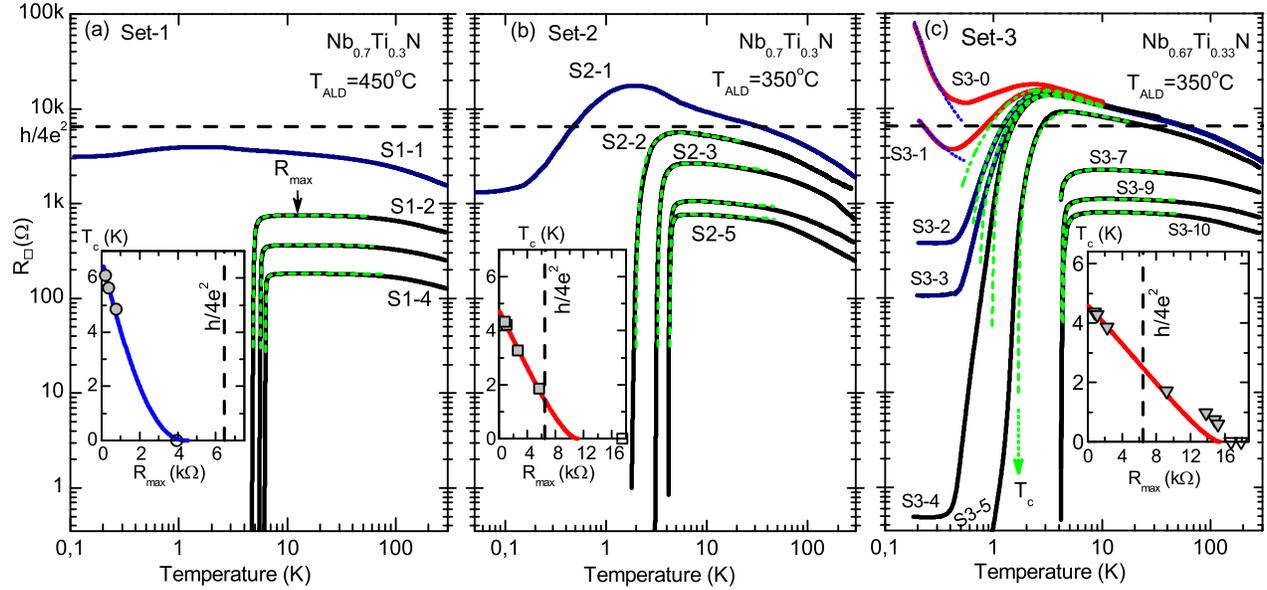}
\vspace{-0.5cm}
\caption{ Sheet resistance $R_{\Box}$ vs. temperature on log-log scale for films of Set-1\,(a), Set-2\,(b) and Set-3\,(c). The deposition temperature $T_\mathrm{ALD}$ and  Ti fraction $x$ in Nb$_{1-x}$Ti$_x$N composition are given on the plots (see SI for the line-log scale).
        The vertical axis scale is same for all plots.
        Solid lines are experimental dependencies.   Horizontal line shows the resistance $R_\mathrm{q} =h/4e^2=6,45$\,k$\Omega$.
        Dashed line: fits accounting for contributions to conductance from superconducting fluctuations (SF), the $T_\mathrm{c}$ obtained from these fits are given in Table.
       (c) Dotted line: activation dependence $R_{\Box} \propto \exp (1/T)$. Sample S3-5 doesn't manifest BKT transition (see Fig.~2 (b), (d)).
     Insets:  $T_\mathrm{c}$ vs. sheet resistance in maximum $R_\mathrm{max}$ prior to superconducting resistance drop, symbols are the experimental values, the solid line is the theoretical fitting by Eq.~(\ref{Fin}) with the adjustable parameter $\gamma =$6.5 ($T_\mathrm{c0}=6.71$\,K) for (a); $\gamma=$4.4 ($T_\mathrm{c0}=4.75$\,K) for (b); and $\gamma=$3.8 ($T_\mathrm{c0}=4.6$\,K) for (c).
     Vertical  line shows the $R_\mathrm{q}$. Open symbols with $T_\mathrm{c}=0$ correspond to samples for which we can not reliably define $T_\mathrm{c}$ with SF-fits.}
    \label{Fig1}
\end{figure*}
%%%%%%%%%%%%%%%%%%%%%%%%%%%%%%%%%%%%%%%%%%%%%%%%%%%%%%%%%%%%%%%%%%%%%%
\section*{Results and discussion}
Figure 1 presents the temperature dependencies
of the sheet resistance $R_{\Box}$ for three our Sets of films.
%where inside one Set films differ in the thickness $d$ and, hence, in the $R_{\Box}$.
Before going into details, we'd like to highlight a qualitative difference between  Set-1 and Sets-2,3. The superconductivity in Set-1 (Fig.~\ref{Fig1}(a)) gets fully suppressed by disorder before the sheet resistance of samples in maximum  $R_\mathrm{max}$ reaches critical value
$R_\mathrm{q} =h/4e^2 = 6.45$ k$\Omega$. In Set-3 (and most likely in Set-2), samples with $R_\mathrm{max} > R_\mathrm{q}$ still experience  superconducting transition, and Sets-2,3 demonstrate more complicated evolution.

In all samples the resistances first grow with cooling down
from room temperature (see Fig.3 in SI for the discussion of high-temperature behaviour of $R_{\Box}(T)$).
With further cooling, the $R_{\Box}(T)$ dependencies first reach maximum $R_\mathrm{max}$ and after that behave differently in different Sets.
%

%\subsection*{Theoretical background}
%
The transition into a superconducting state  is a two-stage process in thin films:
first,  the finite amplitude of the order parameter forms at the superconducting critical temperature $T_\mathrm{c}$ , second, a global phase-coherent
state establishes at lower temperature $T_\mathrm{BKT}$ of the Berezinskii-Kosterlitz-Thouless transition.
Below we analyze suppression of both these temperatures with disorder.

Taking into account perturbative quantum contributions to conductivity from superconducting fluctuations (SF) at $T>T_\mathrm{c}$ and weak localization~\cite{Maki1968, Thompson1970, Lopes1985, Aslamasov1968,  Altshuler1980, Postolova2015} (see SI for details), we fit the experimental data (dashed green lines in Fig.~1).
These SF-fits, in which critical temperature $T_\mathrm{c}$ enters as the adjustable parameter, yield the macroscopic value of $T_\mathrm{c}$. For all samples, the extracted $T_\mathrm{c}$ (given in Table) is very close to the temperature of the inflection point, i.e. the temperature where $dR/dT$ is maximal~\cite{TcBenfatto}, and lies at the foot of $R_{\Box} (T)$~\cite{EPL}.
All Sets demonstrate the suppression of  $T_\mathrm{c}$ with the increase of sample's  `normal' sheet resistance (insets in Fig.~\ref{Fig1}).
In fermionic scenario, the suppression of $T_\mathrm{c}$ follows celebrated Finkelstein's formula~\cite{Finkelstein}:
\begin{eqnarray}
\vspace{-0.1cm}
\ln \left( \frac{T_\mathrm{c}}{T_\mathrm{c0}} \right) = \frac{1}{|\gamma|} - \frac{1}{\sqrt{2r}} \ln \left( \frac{\gamma - r/4 - \sqrt{r/2}}{\gamma - r/4 + \sqrt{r/2}} \right),
\label{Fin}
\vspace{-0.1cm}
\end{eqnarray}
where  $\gamma = 1/ \ln (kT_\mathrm{c0} \tau / \hbar)$ ($T_\mathrm{c0}$ is superconducting critical temperature of a clean sample) and $r=G_{00}R_{\Box _\mathrm{N}}$, where the choice of `normal state' sheet resistance is uncertain due to strong non-monotonic $R_{\Box}(T)$ dependence. Usually, to analyze the suppression of $T_\mathrm{c}$ with disorder, $T_\mathrm{c}$ is plotted vs. resistance at some  temperature~\cite{TcvsRNbN, TcvsRTiN}. For our samples we plot $T_\mathrm{c}$ vs. $R_\mathrm{max}$, the resistance in the maximum prior to superconducting resistance drop.

\begin{figure*}[b!]
\centering
        \includegraphics[ width=0.5\linewidth]{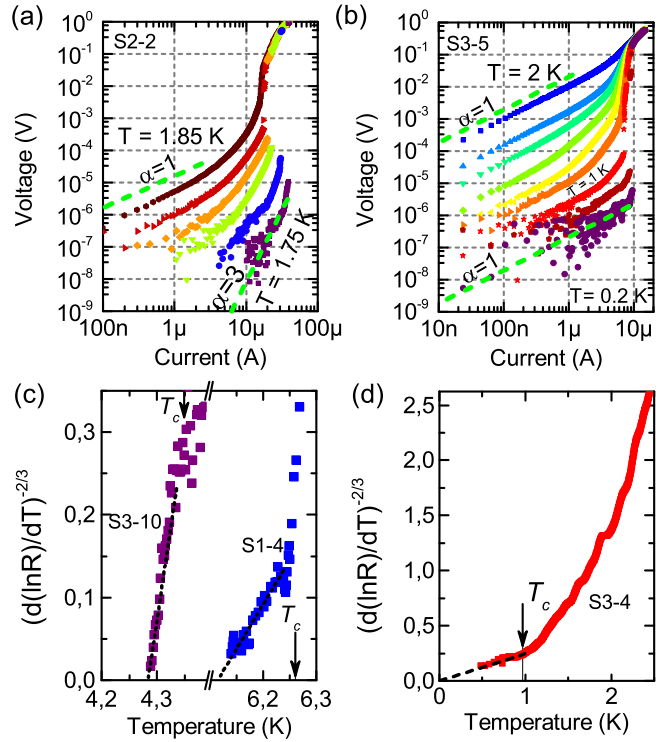}
        \vspace{-0.3cm}
    \caption{
(a), (b) Temperature evolution of current-voltage characteristics on a log-log scale for samples S2-2 and S3-5. Dashed line
        indicate the slopes corresponding to  power $\alpha=1$ and $\alpha=3$ on the $V \propto I^{\alpha}$.
        (c), (d) Rescaling of the sheet resistance $R_{\Box}(T)$  to the BKT form to extract the vortex-unbinding temperature $T_\mathrm{BKT}$ (Eq.~(\ref{BKTfit})) for samples S1-4, S3-10 (c) and sample S3-4 (d), where the straight line (dashed) corresponds to Eq.~(\ref{BKTfit}). Arrows mark position of $T_\mathrm{c}$. Notably, the $R_{\Box}(T)$ curves obey Eq.~(\ref{BKTfit}) just at $T < T_\mathrm{c}$.
        The value $(d(\ln R)/dT)^{-2/3} = 0$ at $T_\mathrm{BKT}$.
        }
    \label{fig2}
\end{figure*}

The BKT transition manifest itself both in the power-law behaviour of current-voltage  characteristics $V \propto I^{\alpha \geq 3}$ and in
exponential decay of resistance:
\vspace{-0.2cm}
\begin{equation}
R(T) \propto \exp (T/T_\mathrm{BKT}-1)^{-1/2}).
\vspace{-0.25cm}
\label{BKTfit}
\end{equation}
The first method for finding $T_\mathrm{BKT}$ is to utilize
power-law fits to the $I$-$V$ characteristics for
which the switch from $V \propto  I$ to $V \propto  I^3$ occurs at the
transition (Fig.~\ref{fig2} (a)).
The second method is to replote $R_(T)$ (see Eq.~(\ref{BKTfit})) as d$(\ln R/$d$T)^{-2/3}(T)$~\cite{BKTderiv}. If the decay of resistance is due to vortex motion the experimental  curve is linear in these coordinates and the intersection of experimental curve with axis X corresponds to $T_\mathrm{BKT}$ (Fig.~\ref{fig2} (c), (d) and Fig.~5 in SI).
As was demonstrated for ALD deposited TiN films~\cite{Postolova2015}, these methods provide BKT temperatures that coincide within one percent.
Below we utilize both methods to observe BKT transition or it's absence in a system.

%%%%%%%%%%%%%%%%%%%%%%%%%%%%%%%%%%%%%%%%%%%%%%%%%%%%%%%%%%%%%%%%%%%%%%
%%%%%%%%%%%%%%%%%%%%%%%%%%%%%%%%%%%%%%%%%%%%%%%%%%%%%%%%%%%%%%%%%%%%%%
\begin{figure}[tbh]
    \begin{center}
        \begin{center}
            \includegraphics[width=0.7\columnwidth]{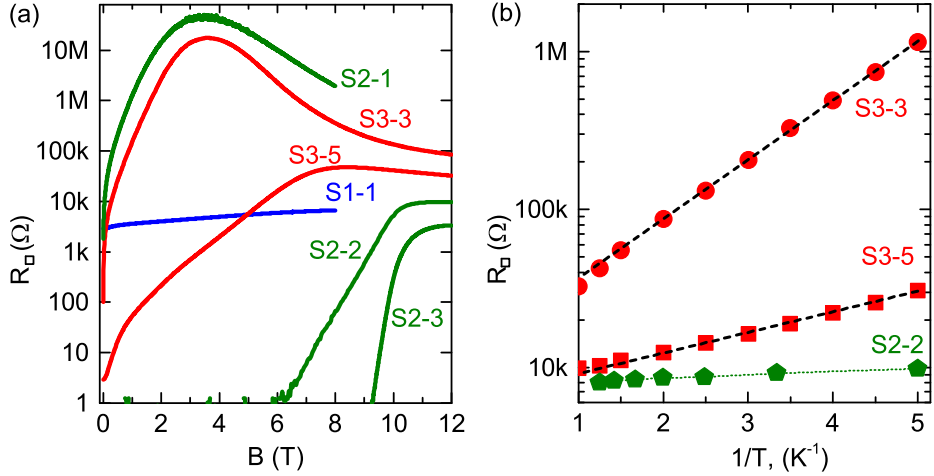}
        \end{center}
        \caption
        {(a) Magnetoresistance per square $R_{\Box} (B)$ on semi-log scale for films listed in figure. All curves are taken at temperature $T=0.2$\,K,
        except for $R_{\Box} (B)$ of sample S1-1 that is obtained at $T=0.04$\,K.
         (b) Arrhenius plot of sheet resistance $R_{\Box}$ in constant perpendicular magnetic field vs. $1/T$ on for samples S3-3 ($B$=1.7\,T), S3-5 ($B$=6.8\,T) and  S2-2 ($B=11.3$\,T). Dashed lines show the activation dependence $R = R_\mathrm{I} \exp (E_\mathrm{I}/k_\mathrm{B}T)$, where for S3-3 $R_\mathrm{I}$=5.3\,k$\Omega$, $E_\mathrm{I}$=86\,meV  and for S3-5
         $R_\mathrm{I}$=2.3\,k$\Omega$, $E_\mathrm{I}$=26\,meV. Sample S2-2 exhibits saturation (not activation).}
        \label{FigRB}
    \end{center}
\end{figure}

\subsection*{Suppression of superconductivity in Set-1}

Figure 1\,(a) demonstrates that the SF-fitting describes fairly well the graduate
decrease in the resistance $R_{\Box}(T)$ matching perfectly the
experimental points down to values $\approx 0.2 \cdot R_\mathrm{max}$ (without
any additional assumptions about mesoscopic inhomogeneities~\cite{BurmBKT}).
Below $T_\mathrm{c}$ the $R_{\Box}(T)$  follows Eq.\,(\ref{BKTfit}) (see Fig.~2 (c) and Fig.~5 in SI) and, hence, is caused by motion of free vortices.
The obtained $T_\mathrm{BKT}$ temperatures are listed in Table.

The values of $T_\mathrm{c}$ plotted vs. $R_\mathrm{max}$ (inset of Fig.~1\,(a))
show that the suppression of $T_\mathrm{c}$ is in accord with  Eq.~(\ref{Fin}), with fitting parameter $\gamma =$6.5. The macroscopic superconductivity in samples is fully suppressed before sample's $R_\mathrm{max}$ reaches $R_\mathrm{q}$.
 Figure~\ref{FigRB} (a) shows $R_{\Box}(B)$ curve of non-superconducting sample S1-1 taken at $T=0.04$\,K. The resistance  depends weakly on magnetic field (without a magnetoresistance peak presented in samples with close  $R_\mathrm{max}$ value from other Sets), implying the lack of Cooper pairs in the sample.

To sum up, the Set-1 exhibits the  superconductor-metal transition in agreement with fermionic scenario.

\subsection*{Suppression of superconductivity in Set-2}

Figure 1\,(b) shows that the graduate resistance drop of $R_{\Box}(T)$ is in agreement with the SF-fitting.
We observe that both the power-law $V(I)$ curves (Fig.~2(a)) and $R(T)$ dependencies at $T<T_\mathrm{c}$  (see Fig.~5 in SI) are in agreement with standard BKT theory
without any additional assumptions about effect of disorder~\cite{BurmBKT} even for sample S2-2 that has $R_\mathrm{max} \lesssim  R_\mathrm{q}$.

The suppression of $T_\mathrm{c}$ in Set-2 (inset in Fig.~1 (b)) is described with the fitting parameter $\gamma=$4.4. That is slightly below the applicability limit of Eq.~(1) which is $\gamma \simeq 5$. Unfortunately, we do not have samples between S2-1 and S2-2 so we can not tell, how $T_\mathrm{c}$ gets suppressed in region $R_\mathrm{max} > R_q$. The sample S2-1 experiences substantial resistance drop, but the global coherent superconducting state is not achieved at lowest temperatures. The behavior of $R_{\Box}(T)$ suggests that the film falls into the Bose metallic state, featuring a finite density of free vortices.

Magnetic field reveals huge difference between $R_{\Box}(B)$ of sample S2-1 ($R_\mathrm{max} > R_\mathrm{q}$)  and samples S2-2, S2-3  ($R_\mathrm{max} < R_\mathrm{q}$)  (Fig.~\ref{FigRB} (a)).
The magnetoresistance of S2-1  shoots up with slight increase of magnetic field from zero, unlike samples with $R_\mathrm{max} < R_\mathrm{q}$ where resistance appears in strong magnetic field and increases relatively slow. Then  $R_{\Box}(B)$ for S2-1 reaches a maximum, followed by decrease of $R_{\Box}(B)$.
We do not see maximum for S2-2 and S2-3  since for superconducting samples it appears at higher fields~\cite{Baturina2007}. As was shown in details in~\cite{Mironov2018}, in constant magnetic field sample S2-1 demonstrates the temperature-driven charge BKT transition into  superinsulating state~\cite{Nature, Mironov2018} --- the state also found in the InO films~\cite{ShaharPRB, Ovadia2015}.

To sum up, the behaviour of Set-2 suggests the action of both fermionic and bosonic mechanisms of superconductivity suppression.

\subsection*{Suppression of superconductivity in Set-3}

The films of Set-3 fall into two categories depending on the ratio between $R_\mathrm{max}$ and $R_\mathrm{q}$ (Fig.~\ref{Fig1}\,(c)).
For  films with $R_\mathrm{max} < R_\mathrm{q}$  the $R_{\Box}(T)$ decrease with cooling down  is in agreement with the conventional theories of superconducting fluctuations  and Berezinskii-Kosterlitz-Thouless transition (Fig.~\ref{fig2}\,(c)), where both  $T_\mathrm{BKT}$ and $T_\mathrm{c}$ decrease with $R_\mathrm{max}$ increasing (see Table).
The  films with $R_\mathrm{max} > R_\mathrm{q}$  show the significant resistance drop (from few orders of magnitude for sample S3-5 to one order for S3-1) the substantial  part of which is in agreement with the SF-fits, implying there are short living Cooper pairs in a system. After deviating from the SF-fits, the $R_{\Box}(T)$ either saturates (samples from S3-5 to S3-2)
 like a Bose metal or increases exponentially (S3-1, S3-0) like an insulator (thinner films, not shown in Fig.~\ref{Fig1} (c), are insulators~\cite{Burdastyh2017}).

The $T_\mathrm{c}$ obtained from the SF-fits decreases with $R_\mathrm{max}$ increasing (inset  in Fig.~\ref{Fig1}\,(c)).
Equation~(\ref{Fin}) describes the $T_\mathrm{c}$ suppression but only for samples with $R_\mathrm{max} < R_\mathrm{q}$ and with quite small $\gamma=$3.8, when normally $\gamma \gtrsim 5$ in fermionic model. For samples with $R_\mathrm{max} > R_\mathrm{q}$ the critical temperature $T_\mathrm{c}$ decreases slower than Eq.~(\ref{Fin}) predicts.
%This may imply that  bosonic mechanism of superconductivity suppression dominants the fermionic one.

Films with $R_\mathrm{max} > R_\mathrm{q}$ do not experience the BKT transition. Particularly, for S3-5 ans S3-4 the $R(T)$ demonstrates decay in accord with Eq.~(\ref{BKTfit}) but yields $T_\mathrm{BKT}=0$ (Fig.~\ref{fig2}(d)) and $V(I)$ curves in low-current limit remain linear $V \propto I ^{\alpha=1}$ at all measured temperatures (Fig.~\ref{fig2} (b)).
This could be due to inhomogeneities of samples, but the estimation of inhomogeneity scale $L$~\cite{Benfatto2009} provides the values $L \simeq 3 - 6$ $\mu$m. This is few orders of magnitude larger that inhomogeneities observed from electronic and atomic-force microscopy.
Even though the exponent $\alpha$ remain $\alpha =1$, the voltage jumps to normal resistance branch with increasing $I$ appear at low temperatures. In disordered ALD deposited TiN films, these voltage jumps in $V(I)$ occur at $T \simeq T_\mathrm{BKT}$, i.e. when $\alpha$ switches from 1 to 3~\cite{IV_jump}. Hence, we observe that the $T_\mathrm{BKT}$ transition vanishes while $T_\mathrm{c}$ remains non-zero.

Representative magnetoresistance curves of samples  with  $R_\mathrm{max} (B=0) > R_\mathrm{q}$ (S3-3 and S3-5) are given in Fig.~\ref{FigRB}\,(a). Both observed features - the magnetoresistance (MR) peak and fast increase of $R_{\Box}(B)$ in weak magnetic field - become more pronounced with sample's $R_\mathrm{max} (B=0)$ increasing.
%Samples exhibit a positive magnetoresistance (PMR) first, then $R(B)$ reaches a maximum, followed by $R(B)$ decrease for S3-3 and S3-5 ($R_\mathrm{max} > R_\mathrm{q}$).
Finally, the behaviour of $R_{\Box}(T, B= \mathrm{const})$ in constant magnetic field (which is slightly smaller that field where $R_{\Box}(B) = \mathrm{max}$) is given in Fig.~\ref{FigRB}\,(b).
Samples with $R_\mathrm{max} (B=0) > R_\mathrm{q}$ demonstrate typical insulating dependence $R = R_\mathrm{I} \exp (E_\mathrm{I}/k_\mathrm{B}T)$.
The giant MR peak and Arrhenius behavior of the resistance
near it may be found in systems in which there are large
fluctuations in the amplitude of the superconducting
order parameter~\cite{Goldman2015}. Experimentally these fluctuation appear to be helped by compositional variations on a
mesoscopic scale~\cite{Goldman2015}. Apparently, the mesoscopic compositional variations emerge in our Sets with decreasing deposition temperature $T_\mathrm{ALD}$ and increasing fraction of Ti $x$.

To sum up here,  for superconducting films with $R_\mathrm{max} > R_\mathrm{q}$, the $T_\mathrm{BKT}$ transition vanishes while $T_\mathrm{c}$ remains non-zero. This is typical for bosonic scenario of suppression of superconductivity~\cite{Gold1986, Fisher1990}. Bearing in mind that $T_\mathrm{c}$  gets suppressed with disorder we conclude that for Set-3 both bosonic and fermionic mechanisms suppress superconductivity.

\section*{Conclusion}
We have examined three sets of superconducting disordered thin Nb$_{1-x}$Ti$_x$N films, where the only difference between sets was the fraction of Ti $x$ and/or the temperature of deposition $T_\mathrm{ALD}$.
We showed that, both increase of $x$ and the decrease of  $T_\mathrm{ALD}$, lead to the smooth crossover from fermionic mechanism of superconductivity suppression to the case where both bosonic and fermionic mechanisms are involved.
We show that the ratio between $R_\mathrm{max}$, and $R_\mathrm{q} =h/4e^2$ divides films of all sets into two categories.
For moderately disordered films ($R_\mathrm{max}  < R_\mathrm{q}$) the superconducting transition is in agreement with the conventional theories of superconducting fluctuations and Berezinskii-Kosterlitz-Thouless transition, and  $T_\mathrm{c}$ decreases with disorder in accord with the Finkel'stein formula. For critically disordered films ($R_\mathrm{max} > R_\mathrm{q}$) the $T_\mathrm{c}$ decreases slower then the Finkel'stein formula predicts.
Moreover, films with $R_\mathrm{max} > R_\mathrm{q}$ do not experience the BKT transition to zero-resistance state.
Careful magnetoresistance measurements revealed that there is a qualitative difference between films with $R_\mathrm{max}$ smaller and greater than $R_\mathrm{q}$.

\section*{Methods}
The fabrication is built upon the Atomic Layer Deposition technique. The structure of films grown on Si substrates with AlN buffer layers was investigated using a JEOL-4000EX electron microscope operated at 400 kV, with a point-to-point resolution of 0.16nm and a line resolution of
0.1 nm. %The details of sample fabrication, analysis and measurements are given in SM.

\section*{Measurement technique}
The films were lithographically patterned into bridges
50 $\mu$m wide, the distance between current-contacts was 2500 $\mu$m and distance between voltage-contacts was 450 $\mu$m.
Low resistive transport measurements ($R(B,T)<1$\,M$\Omega$) are carried out using low-frequency ac and dc techniques in a
four-probe configuration $I = 1-10$\,nA, $f \approx 3$ Hz. High resistive transport measurements ($R(B,T) > 1$\,M$\Omega$) are carried out using low-frequency ac and dc techniques in a two-probe configuration with $V \approx 100$ $\mu$V, $f \approx 1$ Hz. For ac measurements we use one/two SR830 Lock-ins and current/voltage preamplifiers SR570/SR560. For dc measurements we use sub-femtoampermeter Keythley 6430a and nanovoltmeter Agilent 34420.  All resistance measurement are carried out in linear regime with using adequately system of filtration. Resistivity measurements at sub-Kelvin temperatures  were performed  in  dilution refrigerators $^3$He/$^4$He with superconducting magnet.

\section*{Table}
\begin{table}[h!]
\caption{$T_\mathrm{ALD}$ is the deposition temperature; $d$ is film thickness;  $R_\mathrm{max}$ is the resistance at the maximum of $R(T)$; $R_{77}$ is the resistance per square at $T=77$\,K; $T_\mathrm{c}$ is the critical temperature determined from the SF-fits; $T_\mathrm{BKT}$  is BKT transition temperature; $D$ is the diffusion coefficient $D=0.882 \cdot T_c/(e B_\mathrm{c2})$ (see Fig.~2 in SI for $B_\mathrm{c2}$); $n$ is the Hall carrier density (see SI).}
\begin{center}
\begin{tabular}{|c|c|c|c|c|c|c|c|c|}
\hline%------------------------------------------------------------
 \multicolumn{2}{|c|}{sample}           & $d$ &$R_\mathrm{max}$  & $R_{77}$  & $T_\mathrm{c}$& $T_\mathrm{BKT}$& $D$                  & $n$\\
 \multicolumn{2}{|c|}{Nb$_{1-x}$Ti$_x$N}& (nm)&(k$\Omega$)&(k$\Omega$)& (K)    & (K)      & $\frac{\rs{cm}^2}{s}$& $\frac{10^{22}}{\rs{cm}^3}$\\
 \hline%-----------------------------------------------------------
 \rule{0mm}{5mm}
 Set-1, $x=0.3$ & S1-1  & 3   & 3.96      & 2.56      & 0 & 0 & &\\
 \multirow{2}{*}{$T_\mathrm{ALD}$=450$^{\circ}$C}
 &                       S1-2 & 10  & 0.75      & 0.69      & 4.85 & 4.79 & &\\
%\rule{0mm}{5mm}
 &                       S1-4 & 20  &0.18       & 0.17      & 6.26 & 6.11 &  &\\
\hline%------------------------------------------------------------
\rule{0mm}{5mm}
Set-2, $x$=0.3 & S2-1 & 10  & 17.55     &4.52       & 0   & 0 &   & 0.5 \\
\multirow{2}{*}{$T_\mathrm{ALD}$=350$^{\circ}$C}
                       & S2-3 & 15  & 2.66      & 1.69      & 3.27 & 3.08 &  0.2  &\\
%\rule{0mm}{5mm}
                       & S2-5 & 40  & 0.52      &0.77       & 4.33 & 4.18 &   &\\
\hline%------------------------------------------------------------
\rule{0mm}{5mm}
Set-3, $x=0.33$ & S3-1 & 9.2 & 17.9     & ---       & --- & 0   & &\\
%\rule{0mm}{5mm}
\multirow{2}{*}{$T_\mathrm{ALD}$=350$^{\circ}$C}
                       & S3-2 & 9.2  &15.72       & ---      & ---   &  0 &   &\\
                       & S3-3 & 9.2   & 15.18     & 5.75      & 0.75   & 0   & &\\
                       & S3-4 & 9.2  & 14.13        & ---      & 0.97  & 0 &   &\\
                       & S3-5 & 10  &9.26       & 4.53      & 1.7  & 0 &   &\\
%\rule{0mm}{5mm}
                       & S3-7 & 12  & 2.24      & 1.87      & 3.85 & 3.81 &  &\\
%\rule{0mm}{5mm}
                       & S3-9 & 19  & 1.87      & 0.98      & 4.28 & 4.26 &  0.3  & 1 \\
%\rule{0mm}{5mm}
                       & S3-10& 21  & 0.8       & 0.69      & 4.35 & 4.29  &  &\\
\hline%------------------------------------------------------------
\end{tabular}
\end{center}
\end{table}

%%%%%%%%%%%%%%%%%%%%%%%%%%%%%%%%%%%%%%%%%%%%%%%%%%%

%\bibliography{dynamicalmott}

%%%%%%%%%%%%%%%%%%%%%%%%%%%%%%%%%%%%%%%%%%%%%%%%%%%

\section*{Acknowledgements}

The experimental work was supported by grant of RF president (MK-5455.2018.2).
The work at Argonne (V.M.V. and T.P.) was supported by the
U.S. Department of Energy, Office of Science, Basic
Energy Sciences, Materials Sciences and Engineering
Division. The Hall and magnetoresistance measurements of sample S3-9 performed in IPM PAS was supported by RSF, project No. 15-12-10020. We are grateful to Dr. Tatyana I. Baturina  for initiating the work on NbTiN films.

\section*{Author contributions statement}

The films were synthesized by TP; AYuM, SVP, MVB carried out the experiments;  SVP, SSU, AVA carried out magnetoresistance measurements of sample S3-9;
AYuM, SVP and VMV analyzed the data.
All authors discussed the results and contributed in writing the manuscript.

\section*{Additional information}
The authors declare that they have no competing interests as defined by Nature Research, or other interests that might be perceived to influence the results and/or discussion reported in this paper.

\section*{Data availability}
The authors declare that all relevant data supporting the findings of this study are available within the article and its supplementary information file. Additional raw data, if necessary, are available upon request to AYuM, mironov@isp.nsc.ru

\end{document}